\documentclass[a4paper,11pt]{article}

\usepackage{bm}
\usepackage{latexsym}
\usepackage{amssymb}
\usepackage{amsmath}
\usepackage{mathtools}
\usepackage{hyperref}
\hypersetup{colorlinks=true, allcolors=blue}
\usepackage{graphicx}

\usepackage{mathbbol}
\usepackage{braket}
\usepackage{empheq}
\usepackage{cite}
\usepackage[title]{appendix}
\usepackage{here}
\usepackage{pifont}
\usepackage{comment}
\usepackage[normalem]{ulem}

\makeatletter
\renewcommand\subparagraph{
    \@startsection {subparagraph}{5}{\z@ }{3.25ex \@plus 1ex
    \@minus .2ex}{-1em}{\normalfont \normalsize \bfseries }}
\makeatother

\numberwithin{equation}{section}

\begin{document}
\pagestyle{empty}

\vspace{-4cm}
\begin{center}
    \hfill KEK-TH-2787, KOBE-COSMO-25-18, YITP-25-190 \\
\end{center}

\vspace{2cm}

\begin{center}

{\bf\LARGE
Quantum sensing of high-frequency gravitational waves with ion crystals \\
}

\vspace*{1.5cm}
{\large 
Asuka Ito$^{1}$, Ryuichiro Kitano$^2$, Wakutaka Nakano$^2$
and Ryoto Takai$^{2,3,4}$
} \\
\vspace*{0.5cm}

{\it 
$^1$Department of Physics, Kobe University, Kobe 657-8501, Japan \\
$^2$Yukawa Institute for Theoretical Physics, Kyoto University,
Kyoto 606-8502, Japan \\
$^3$KEK Theory Center, Tsukuba 305-0801, Japan \\
$^4$The Graduate University for Advanced Studies (SOKENDAI), Tsukuba
305-0801, Japan\\
}

\end{center}

\vspace*{1.0cm}

\begin{abstract}
{\normalsize \noindent
A detection method for high-frequency gravitational waves using two-dimensional ion crystals is investigated.
Gravitational waves can resonantly excite the drumhead modes of the ion crystal, particularly the parity-odd modes.
In the optical dipole force protocol, entanglement between the drumhead modes and the collective spins transfers the excitation of the drumhead modes to the rotation of the total spin. 
Furthermore, gravitational wave detection beyond the standard quantum limit becomes possible as a squeezed spin state is generated through this entanglement.
The sensitivity gets better with a larger ions crystals as well as a larger number of the ions.
Future realization of large ion crystals can significantly improve the sensitivity to  
gravitational waves in the 10~kHz to 10~MHz region.
}
\end{abstract} 


\newpage
\baselineskip=18pt
\setcounter{page}{2}
\pagestyle{plain}

\setcounter{footnote}{0}

\tableofcontents
\noindent\hrulefill


\section{Introduction}

Gravitational wave detection with LIGO has opened the era of gravitational wave 
astronomy/cosmology\cite{LIGOScientific:2016aoc,LIGOScientific:2016emj,LIGOScientific:2016vbw,LIGOScientific:2016vlm,LIGOScientific:2017vwq}.
Gravitational wave interferometers have detected gravitational waves of around kHz from binaries of black holes or neutron stars, allowing us to probe new physics, such as testing general relativity~\cite{LIGOScientific:2021sio} 
and exploring the astrophysical history of the 
universe~\cite{Fowler:1964zz,Barkat:1967zz,Woosley:2021xba,Sasaki:2016jop}.
Recently, pulsar timing arrays have reported the detection of gravitational waves~\cite{NANOGrav:2023gor,EPTA:2023fyk,Xu:2023wog,Reardon:2023gzh}, whose origin is considered to be supermassive black hole binaries~\cite{Ellis:2023dgf} 
or the decay of topological defects such as cosmic strings~\cite{Ellis:2020ena}.
Obviously, advancing multi-frequency gravitational wave observations is crucial to exploring our 
universe~\cite{Kuroda:2015owv}. 
The detection of gravitational waves at frequencies higher than kHz is theoretically interesting for probing new physics~\cite{Aggarwal:2025noe}. However, it is still under development, and novel ideas are desired~\cite{Ito:2019wcb,Ito:2020wxi,Ejlli:2019bqj,Berlin:2021txa,Domcke:2022rgu,Tobar:2022pie,Berlin:2023grv,Ito:2022rxn,Pshirkov:2009sf,Dolgov:2012be,Domcke:2020yzq,Ramazanov:2023nxz,Ito:2023fcr,Takai:2025cyy,Ito:2023bnu,Ito:2023nkq,Matsuo:2025blj}. 
Since experimental sensitivity is maximized when the wavelength of target gravitational waves is comparable to the detector size, one natural experimental scheme for detecting high-frequency gravitational waves is to employ table-top experiments.
In particular, quantum sensing can be a promising way to improve sensitivity~\cite{Ito:2023bnu}.

Quantum technologies have advanced toward the realization of quantum computers.
For instance, laser technology enables us to trap ions stably and maintain entangled states for extended periods.
This apparatus can also be used as detectors for probing weak electric or magnetic 
fields~\cite{Gilmore:2017cbw,Affolter:2020jbe}. 
In particular, quantum sensing with the use of quantum entanglement or squeezed states improves the sensitivity compared to conventional methods and can approach the Heisenberg 
limit~\cite{Giovannetti:2011chh,Pezze:2016nxl,Gilmore:2021qqo}. 
Quantum sensing is also useful for searching for physics beyond the Standard Model, such as dark matter~\cite{Gilmore:2021qqo,Fan:2022uwu,Chen:2022quj,Afek:2021vjy,Carney:2021irt,Budker:2021quh,Berlin:2022hfx,Brady:2022bus,Brady:2022qne,Shi:2022wpf,Sushkov:2023fjw,Ikeda:2021mlv,Chigusa:2023roq,Chigusa:2023szl,Chen:2023swh,Ito:2023zhp,Chen:2024aya,Shu:2024nmc,He:2024sfi}.

For specific detectors, increasing the scale of the
detector can indeed improve detection sensitivity, thereby discovering more
signals.
In particular, it has been discussed in Ref.~\cite{Ito:2023bnu}
that high-frequency gravitational waves can be detected by using 
a single electron in a Penning trap system.
Unlike the sensing of the electric field, the gravitational wave signal 
is enhanced when the size of the wave function (or the magnetron orbits) of the trapped 
electron is large.
This observation motivates us to consider an even larger system,
such as ion crystals in Penning trap systems where
a large number of ions are aligned on the two-dimensional plane.

In this paper, we investigate a detection method for high-frequency gravitational waves using two-dimensional ion crystals~\cite{britton2012engineered,sawyer2012spectroscopy}.
The ion crystal is formed in the Penning trap through a balance between the trap potential and Coulomb repulsion among the ions
and exhibits collective oscillations known as drumhead modes. 
Ref.~\cite{Gilmore:2021qqo} has demonstrated an effective use of the ion crystal with about 150 $^9$Be$^+$ ions as a sensor to weak electric fields.
We show that parity-odd modes can be excited by gravitational waves, and parity-even modes are suppressed due to the quadrupole nature of gravitational waves. 
This feature provides a unique signature of gravitational waves, which is useful to discriminate them from other sources.
The excitation of the axial oscillation is read out as a rotation of the collective spin with the optical dipole force, enabling sensitivity beyond the standard quantum limit via squeezed spin states~\cite{leibfried2003experimental,Giovannetti:2011chh,
britton2012engineered,sawyer2012spectroscopy}. 


\section{Experimental setup for detecting odd mode excitation}

In this section, we discuss a detection scheme for external fields,
such as gravitational waves, using the odd modes of a spin-squeezed ion crystal,
which allows sensitivities beyond the standard quantum limit.
The excitation of odd modes would be characteristic of gravitational waves,
making the scheme essential for distinguishing them from other sources.
Below, we basically follow the setup of the electric field sensor realized
by the Penning trap system in Ref.~\cite{Gilmore:2021qqo}.
The $^9{\rm Be}^+$ ions are aligned on a two-dimensional plane, and its vibrational mode
is coupled to a spin direction of the outer most orbit electron by using the 
optical dipole force described below. The spin rotation can be used as the 
detection of the external force.
We here discuss a modification of the system suitable for the detection of 
high-frequency gravitational waves.


\subsection{Ion crystals}
\label{sec:odd_ioncrystal}

We briefly review the setup of two-dimensional ion crystals in this subsection.
The ions are confined by a quadrupole electric field and a
magnetic field in a Penning trap. 
Moreover, they rotate in the same phase as if a rigid body around the center of the crystal 
with the rotating wall electric field~\cite{PhysRevLett.80.73,huang1998precise,Gilmore:2021qqo}.

In the frame rotating with ions, the effective potential of a single ion is given
as the harmonic oscillators in the axial and radial
directions~\cite{huang1998precise}, 
\begin{equation}
    V = \frac{1}{2} m_{\rm ion} \omega_z^2 \big( z^2 + \beta \, r^2 \big),
    \label{eq:potential}
\end{equation}
with $m_{\rm ion}$ and $\omega_z$ denoting the ion mass and the frequency of
the axial mode, respectively.
Here, we take the direction of the magnetic field as the $z$-axis.
This system is stable if the radial confinement strength $\beta = \omega_r
(\omega_c - \omega_r) / \omega_z^2 - 1/2$ is positive, where
$\omega_r$ is the phase-locked rotation frequency of the ions, and $\omega_c
= e B / m_{\rm ion}$ is the cyclotron frequency with the magnetic field
$B \sim 1$~T.

Multiple ions in the trap form a two-dimensional Wigner crystal, whose
equilibrium points are generated by the triangular lattice vectors
${\bm a}_1 = (1, 0, 0)$ and ${\bm a}_2 = (1/2, \sqrt{3}/2,
0)$~\cite{porras2006quantum}.
The potential energy of the axial motion is given by
\begin{align}
    V_z &= \frac{1}{2} m_{\rm ion} \omega_z^2 \sum_{i=1}^N z_i^2 +
    \frac{1}{2} \sum_{i \neq j} \frac{\alpha_{\rm EM}} {\sqrt{\lvert
    {\bm \rho}_i - {\bm \rho}_j \rvert^2 + (z_i - z_j)^2}}  \\
    &= \frac{1}{2} m_{\rm ion} \omega_z^2 \sum_i z_i^2 -
    \frac{\alpha_{\rm EM}}{4} \sum_{i \neq j} \frac{(z_i - z_j)^2}{\lvert
    {\bm \rho}_i - {\bm \rho}_j \rvert^3} + \cdots, \label{pote}
\end{align}
where $\alpha_{\rm EM} = e^2 / 4 \pi \simeq 1/137$ is the fine-structure
constant, $z_i$ is the $z$ component of the $i$th ion position, and
${\bm \rho}_i = a_0 (\rho^1_i {\bm a}_1 + \rho^2_i {\bm a}_2)$ with two integers
$\rho^1_i$, $\rho^2_i$ and the lattice constant $a_0 \sim 10$~µm is its
equilibrium position in the crystal plane.

The ions which behave approximately as quantum harmonic oscillators couple
with each other by Coulomb forces, thereby constructing $N$ collective phonon
modes along the $z$-axis.
The Hamiltonian in the harmonic approximation is written as $H_0 =
\sum_k \omega_k \, \hat{a}^\dagger_k \, \hat{a}_k$ by expanding $\hat{z}_i$ by the ladder
operators of eigenmodes $\hat{a}_k^{\dagger}$ at eigenfrequency
$\omega_k$ as 
\begin{equation}
    \hat{z}_i = \sum_k \frac{b_{ik}}{\sqrt{2 m_{\rm ion} \omega_k}} \left(
    \hat{a}_k e^{-i \omega_k t} + \hat{a}^\dagger_k e^{i \omega_k t} \right),
    \label{eq:zmode}
\end{equation}
with the orthogonal matrix $b_{ik}$ diagonalizing the potential $V_z$.
It exhibits drumhead modes in the large ion number limit, as discussed in Appx.~\ref{sec:drumhead}.
The center-of-mass mode, in which all ions move in the same phase, 
has the highest frequency, $\omega_1 = \omega_z$, and
the eigenvector is given by $b_{i1} = 1 / \sqrt{N}$ for all $i$. 
The two axially asymmetric and parity-odd modes have the succeeding frequencies $\omega_2 = \omega_3$,
though this degeneracy may be lifted due to the distortion of the
boundary of the ion cloud by the rotating wall potential~\cite{sawyer2012spectroscopy}.

Henceforth, we specialize our attention to the $^9$Be$^+$ ion, although
the following discussion would apply to some other ion species.
The $^{9}$Be$^{+}$ ion has a nuclear spin of $I = 3/2$ and is optically pumped to 
the $m_I = +3/2$ state in our experimental setup.
The single valance electron is installed as the qubit $\Ket{\uparrow}
\equiv \Ket{m_J = +1/2}$ and $\Ket{\downarrow} \equiv \Ket{m_J = -1/2}$
in the $S_{1/2}$ manifold, with the frequency gap being $\sim 100$~GHz for
$B \sim 1$~T.
A microwave at the frequency is implemented for the global rotation of the
spins and makes the transition between $\Ket{\uparrow}$ and
$\Ket{\downarrow}$ in $\sim 10$~µs. 
Doppler cooling and state preparation are performed by the transitions
via the $P_{3/2}$ states~\cite{Gilmore:2021qqo}.
The state initialization and readout can be performed with high
fidelity~\cite{biercuk2009optimized}. 

\subsection{Optical dipole force}
\label{sec:odd_odf}

The optical dipole force (ODF) produces entangled states between the collective spin of the crystal and 
the phonon modes, which provides opportunities to simulate an Ising-type spin-spin interaction with
various power-law scalings~\cite{britton2012engineered}. 
The excitation of the phonon modes can be detected by observing the spin states, 
enabling high-sensitivity electric field sensing~\cite{Gilmore:2017cbw,Affolter:2020jbe},
even beyond the standard quantum limit~\cite{Gilmore:2021qqo}.

The ODF is implemented by employing two lasers applied to the ion crystal from above and below the plane, 
namely,
${\bm E}_a ({\bm x}, t) \propto {\bm \epsilon}_a \cos ({\bm k}_a \cdot {\bm x} - \omega_a t)$
($a = {\rm U, \, L}$), where ${\bm \epsilon}_a$, ${\bm k}_a$ and
$\omega_a$ are the polarization vector, wavenumber vector, and
angular frequency, respectively. 
The upper beam is set as ${\bm k}_{\rm U} = k (0, - \cos (\theta_0 / 2),
- \sin (\theta_0 / 2))$ and ${\bm \epsilon}_{\rm U} = (\sin \phi_0, 0,
\cos \phi_0)$, and the lower beam as ${\bm k}_{\rm L} = k (0, - \cos (\theta_0 / 2), \sin (\theta_0 /
2))$ and ${\bm \epsilon}_{\rm L} = (- \sin \phi_0, 0, \cos \phi_0)$,
satisfying $0 < \theta_0 < \phi_0 < \pi / 2$.

Each laser individually induces the AC Stark shifts
\begin{equation}
    \Delta_{\kappa} = \Delta^\parallel_{\kappa} \cos^2\phi_0 
    + \Delta^\perp_{\kappa} \sin^2\phi_0 ,
\end{equation}
for the states $\Ket{\kappa} = \Ket{\uparrow}$ and $\Ket{\downarrow}$,
where $\Delta^\parallel_{\kappa}$ and $\Delta^\perp_{\kappa}$ denote the energy shifts of the state $\Ket{\kappa}$ for the polarization components
parallel and perpendicular to the $z$-axis, respectively. 
The shift of the transition energy between the $\Ket{\uparrow}$ and $\Ket{\downarrow}$ states is given by 
$\Delta_{\uparrow} - \Delta_{\downarrow}$,
which is canceled by choosing an appropriate value of the polarization angle
$\phi_0$. 
Consequently, only the interference term of the two lasers, $2 {\bm E}_{\rm U}\cdot {\bm E}_{\rm L}$,
yields the spatially dependent AC Stark shift
\begin{equation}
    \tilde{\Delta}_\kappa = 2 \left( \Delta^\parallel_{\kappa} \cos^2 \phi_0 
    - \Delta^\perp_{\kappa} \sin^2 \phi_0 \right)
    \cos (\delta k \, z - \omega_{\rm ODF} \, t),
\end{equation}
with $\delta k = \lvert {\bm k}_{\rm U} - {\bm k}_L \rvert = 2 k \sin (\theta_0 / 2)$
and $\omega_{\rm ODF} = \vert \omega_{\rm U} - \omega_{\rm L} \vert$.
Therefore, we obtain the state-dependent force
\begin{equation}
    F_{\kappa}(z,t) \simeq F_{0,\kappa} \, z \cos(\omega_{\rm ODF} \, t).
\end{equation}
Here, we assume $\vert \delta k\, z \vert \ll 1$ and define
$F_{0,\kappa} = -2\delta k \left( \Delta^\parallel_{\kappa} \cos^2 \phi_0  
- \Delta^\perp_{\kappa} \sin^2 \phi_0 \right)$.
One can also choose the detuning of the two ODF lasers to have
$F_{0,\uparrow} = - F_{0,\downarrow} \equiv F_0$.

We then obtain the spin-dependent force $F_i = F_{0} \, z_i \sigma^z_i \cos (\omega_{\rm
ODF} \, t)$ for each ion $i$,
with $\sigma^z_i$ denoting the third Pauli matrix.
The Hamiltonian of the optical dipole force is 
\begin{equation}
    H_{\rm ODF} = \sum_i F_0 \cos(\omega_{\rm ODF} \, t) \, z_i \, \sigma^z_i.
    \label{eq:odf}
\end{equation}
This homogeneous ODF is effective in detecting the displacement
of the center-of-mass mode~\cite{Gilmore:2021qqo}. 
However, it provides little sensitivity to other modes,
since the linear signal contributions
cancel out in the protocol explained in the next subsection.

One may be able to prepare an inhomogeneous ODF to selectively couple the
ODF to desired modes by using deformation mirrors~\cite{Polloreno:2022nxl}.
The phase of the ODF laser beams can be controlled, which allows
the cosine factor in Eq.~\eqref{eq:odf} to depend on the ion positions
as $\cos(\omega_{{\rm ODF}} \, t + f(\rho_i) \cos \phi_i)$.
Here, $f(\rho_i)$ represents a function of $\rho_i$, which is the radial distance of the $i$th ion from the center of the ion crystal 
in the $xy$-plane, and $\phi_i$ is the azimuthal angle.
When one adjusts the ODF frequency $\omega_{{\rm ODF}}$ to a target phonon mode $\omega_2$ and takes a proper function $f(\rho_i)$,
the ODF Hamiltonian is obtained as
\begin{equation}
    \hat{H}_{\rm ODF} \simeq \frac{g}{\sqrt{N}} \left(
    \hat{a} + \hat{a}^\dagger \right) \hat{J}_z,
    \quad \hat{a} = \frac{1}{\sqrt{2}} \left(
    \hat{a}_2 + i \hat{a}_3 \right) ,
    \label{eq:odf2}
\end{equation}
where $g$ is the coupling strength, 
denoted as $g_2$ in Appx.~\ref{sec:odf}.
We assume that two modes, $k = 2$, 3, are degenerate within the
frequency detuning of the ODF laser.
See Appx.~\ref{sec:odf} for the derivation of Eq.~\eqref{eq:odf2}.
The phase between the ODF and the phonon modes is synchronized, as can be done experimentally~\cite{Affolter:2020jbe,Gilmore:2021qqo}.


\subsection{Quantum enhanced detection protocol}
\label{sec:odd_protocol}

In the previous subsection, we see that the inhomogeneous ODF would be able to
applied for the parity-odd mode.
It is demonstrated in this subsection that the excitation of the phonon mode can be detected
by measuring the total spin through the ODF, following
the procedure outlined in Ref.~\cite{Gilmore:2021qqo}, 
which enables high-sensitivity detection of external fields including gravitational waves.
We start from the assumption that there is an external wave resonantly coupling to the phonon mode of the ion crystal with the form of
the interaction Hamiltonian
\begin{equation}
    \hat{H}_{\rm int} = \alpha \, \hat{a} +
    \alpha^* \, \hat{a}^\dagger,
    \label{alpha}
\end{equation}
where $\alpha$ represents the amplitude of the target wave.

\begin{figure}[t]
    \centering    
    \includegraphics[width=1\textwidth]{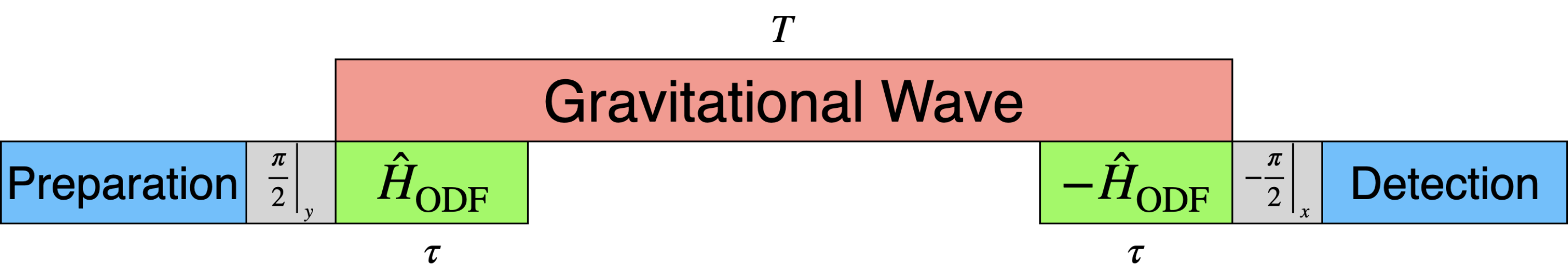}  
\caption{
Schematic illustration of the Ramsey-type experimental sequence 
for the detection of external fields, such as gravitational waves.
The sequence consists of state preparation, an ODF pulse of duration $\tau$, 
free evolution under a gravitational wave for a time $T$, an inverse ODF pulse, 
and final spin detection.
}
    \label{fig:sequence}
\end{figure}

We initially prepare the state 
$\Ket{{\rm init}} = \Ket{N/2}_z \otimes  \Ket{0}$, where
$\Ket{M}_a$ represents the eigenstate of the total spin
operator $\hat{J}_a = \sum_i \hat{\sigma}^a_i / 2$, and 
$\Ket{0}$ is the vacuum of $\hat{a}$.
We use a scheme similar to the Ramsey interferometer, i.e.,
the spin rotation operator $U_y (\pi / 2) = \exp (- i
\frac{\pi}{2} \hat{J}_y)$ is applied to $\Ket{{\rm init}}$ first.
Next, the ODF is performed for the time $\tau$ to change the sensor
state to $e^{-i \tau (\hat{H}_{\rm ODF} + \hat{H}_{\rm int})} \, U_y (\pi / 2) \Ket{{\rm init}}$. 
Afterward, we wait for the time $T - 2 \tau$. The time evolution of the state
is described by the displacement operator $e^{-i (T - 2 \tau) \hat{H}_{\rm int}}$.
The inverse of the ODF operation, $- \hat{H}_{\rm ODF}$, is performed.
In practice, the inverse operator is equipped by the combination of
the ODF $\hat{H}_{\rm ODF}$ and the spin-flip operator $U_x (\pi)$.
Since the spin-flip operator commutes with the displacement operator,
it can be implemented within the waiting period.
Finally, after a $\pi / 2$ pulse around the $x$-axis, $U_x (\pi/2)$, we obtain the final state as
\begin{align}
    \Ket{{\rm fin}} &\equiv U_x \left( \frac{\pi}{2} \right)
    e^{-i \tau (\hat{H}_{\rm ODF} + \hat{H}_{\rm int})} \, U_x (\pi)
    e^{-i (T - 2 \tau) \hat{H}_{\rm int}} \, e^{-i \tau
    (\hat{H}_{\rm ODF} + \hat{H}_{\rm int})} \, U_y \left( \frac{\pi}{2}
    \right) \Ket{{\rm init}} \\
    &= U_y (-\varphi) \Ket{\frac{N}{2}}_{\!\! x} \otimes
    e^{-i T \hat{H}_{\rm int}} \ket{0} . \label{ent}
\end{align}
with
\begin{equation}
    \varphi = \frac{2 g \tau (T - \tau)}{\sqrt{N}} \,
    ({\rm Im} \, \alpha).  \label{phi}
\end{equation}
The experimental sequence of operations is illustrated in Fig.~\ref{fig:sequence}.

Although we start from the ground state of the phonon mode for simplicity,
the evolution of the spin state is independent of the choice.
From Eq.~\eqref{phi}, the signal $\varphi$ can be maximized by controlling the phase of the ODF
if one knows the phase of the external waves.
Otherwise, ${\rm Im} \, \alpha$ would be observed as 
an averaged value in many observation trials.

The signal $\eta \equiv \vert {\rm Im} \, \alpha \vert$
is encoded in the expectation value of $\hat{J}_z$ as
\begin{equation}
    \left\vert \langle \hat{J}_z \rangle \right\vert = \frac{N}{2} \vert \sin
    \varphi \vert \simeq \sqrt{N} g \tau (T - \tau) \, \eta .
\end{equation}
For ideal measurements without noises, the variance of $\hat{J}_z$ is given
by ${\rm var} (\hat{J}_z) = N / 4 + \mathcal{O} (\eta^2)$.
We define the sensitivity to $\eta$ as $\delta \eta$, satisfying $\eta
/ \delta \eta = {\rm SNR}$ with the signal-to-noise
ratio ${\rm SNR} = \left\vert \langle \hat{J}_z \rangle \right\vert /
\sqrt{{\rm var} (\hat{J}_z)}$
and then obtain $\delta \eta = 1 /( 2 g \tau (T - \tau))$ for a single measurement. 
When the measurement is repeated $N_{\rm rep}$ times, the sensitivity is improved as $N_{\rm rep}^{-1/2}$.
The sensitivity $\delta \eta$ is minimized when the duration of the ODF is chosen as $\tau = T/2$. 
Moreover, sensitivity beyond the standard quantum limit, $\delta \eta = 1/(2T)$, can be achieved.
This improvement is attributed to the fact that the ODF can be regarded as a squeezing operation of the state of a quadrature defined by a linear combination of the spin operators and the phonon ladder operators~\cite{Gilmore:2021qqo}.
This total spin $\langle \hat{J}_z \rangle$ is measured by the spin-dependent
global resonance fluorescence
method~\cite{lewis2020protocol,hosten2016quantum,Gilmore:2021qqo}.
We assume that the error of this readout is negligible for simplicity.

In actual experiments, there are several sources of background events.
In Appx.~\ref{sec:noise}, we evaluate the noise and dissipation effects which determine the
variance of $\hat{J}_z$.
The single-measurement sensitivity is given by 
\begin{equation}
    (\delta \eta)^2 \simeq \frac{e^{2 \Gamma \tau}}{4 g^2 \tau^2 (T-\tau)^2} 
    + \frac{e^{2 \Gamma \tau} \sigma^2 (2T^2 - \tau T + \tau^2)}{8 g^2
    \tau^2 (T-\tau)^2}+ \frac{\sigma^2 (2 \bar{n} + 1)}{4} + \frac{g^2
    \tau^2 \sigma^2 (T-\frac{4}{3}\tau)^2}{4(T-\tau)^2} ,
    \label{eq:odd_eta0}
\end{equation}
where $\sigma^2$ is the variance of the frequency detuning of the ODF laser from the phonon mode frequency,
$\bar{n}$ is the occupation number of the phonon mode, and
$\Gamma$ is the decoherence rate.
Here we assume that the heating noise is small enough to be ignored.
The first term reproduces the ideal result in the limit $\Gamma \tau \to 0$.
From Eq.(\ref{eq:odd_eta0}), we see that 
the sensitivity can go beyond the standard quantum limit by tuning the experimental parameters 
even in the presence of the noise, as discussed in~\cite{Gilmore:2021qqo}.

In the current experiment using the ODF protocol, $\Gamma \sim 250$~s$^{-1}$,
$\sigma \sim 2 \pi \times 40$~Hz, and $\bar{n} \sim 5$ were obtained
with $N \sim 150$ ions~\cite{Gilmore:2021qqo}, which can be improved to
$\sigma \sim 2\pi \times 1$~Hz and $\bar{n} \sim 0.3$,
as demonstrated in another experiment~\cite{jordan2019near}.
Furthermore, increasing the number of ions can enhance the sensitivity.
For instance, an ion number of $N \sim 10^8$ has already been demonstrated 
experimentally~\cite{huang1998precise}, 
although the resulting crystal in this case is three-dimensional.


\section{Gravitational wave coupling and sensitivity}
\label{sec:odd_sensitivity}

One needs to use an appropriate coordinate system to investigate the effect of 
gravitational waves on the ion crystal correctly. 
The most useful coordinate would be the proper detector frame~\cite{Manasse:1963zz,Ni:1978zz}, which is co-moving with the 
crystal in our case.
More specifically, the origin of the coordinate is set to coincide with the center of motion of the crystal which 
is freely falling,
so that the coordinate values properly trace the position of each ion.
In the non-relativistic limit, the interaction between a gravitational field and 
an ion is given by 
$\frac{1}{2} T^{\mu\nu} \delta g_{\mu\nu} \simeq \frac{1}{2} T^{00} \delta g_{00}$ where
$T_{\mu\nu}$ and $\delta g_{\mu\nu}$ represent the energy-momentum tensor of the ion and 
the perturbed spacetime metric, respectively.
In the proper detector frame, the interaction Hamiltonian is
\begin{equation}
    H_{\rm GW} = \frac{1}{2} m_{\rm ion}  R_{0k0l} \, x^k x^l ,
    \label{Hint}
\end{equation}
where $R_{\mu\nu\lambda\sigma}$ is the Riemann tensor evaluated at the origin of the coordinate, $x^{i}=0$.
We note that although gravitational waves can couple to the spin of the ion 
directly, such effects 
are suppressed compared with the above interaction~\cite{Ito:2020wxi,Ito:2020xvp,Ito:2022rxn}.

We now consider gravitational waves on the Minkowski spacetime, the
metric is given by $g_{\mu\nu} = \eta_{\mu\nu} + h_{\mu\nu}$, where
$\eta_{\mu\nu}$ is the Minkowski metric and $h_{\mu\nu}$ is the
traceless transverse metric, namely, gravitational
waves. Then, the Riemann tensor is 
\begin{equation}
    R^{\alpha}{}_{\mu\beta\nu} 
    =  \frac{1}{2}(h^{\alpha}_{\ \nu,\mu\beta}-h_{\mu\nu\ ,\beta}^{\ \ ,\alpha}
    -h^{\alpha}_{\ \beta,\mu\nu}+h_{\mu\beta\ ,\nu}^{\ \ ,\alpha}) \ . \label{Rie}
\end{equation}
Although the transverse-traceless gauge is adopted for computing the components of the Riemann tensor, the result remains consistent with that in the proper detector frame, 
since the Riemann tensor is gauge invariant at the linear order.
For the gravitational wave that is responsible for exciting the phonon modes of the crystal, 
we consider a plane wave,
\begin{equation}
    h_{ij}(\bm{x}, t) = h^{(+)} \cos \left( \omega t -
    \bm{k} \cdot \bm{x} + \phi^{(+)} \right) e^{(+)}_{ij}
    + h^{(\times)} \cos \left( \omega t - \bm{k} \cdot \bm{x}
    + \phi^{(\times)} \right) e^{(\times)}_{ij} ,
    \label{planer}
\end{equation}
where
$\omega = 2 \pi f$ is the angular frequency, ${\bm k}$ is the
wavevector, $h^{(+)}$ and $h^{(\times)}$ are the amplitudes of two
polarizations, and $\phi^{(+)}$ and $\phi^{(\times)}$ denote these phases.
The polarization tensors are explicitly constructed as
\begin{align}
    e _{ij}^{(+)} &= \frac{1}{\sqrt{2}}\left(
    \begin{array}{ccc}
      \cos^2 \theta & 0 & -\cos\theta \sin\theta \\
      0 & -1 & 0 \\
      -\cos\theta \sin\theta & 0 & \sin^2 \theta
    \end{array} 
    \right) , \label{lipo1}  \\
    e _{ij}^{(\times)} &= \frac{1}{\sqrt{2}}\left(
    \begin{array}{ccc}
      0 & \cos\theta & 0 \\
      \cos\theta & 0 & -\sin\theta \\
      0 & -\sin\theta & 0
    \end{array} 
    \right) ,  \label{lipo2}
\end{align}
with the $+$ mode defined as a deformation in the $y$ direction and
$\theta$ being the angle in the $zx$-plane. 
By substituting Eqs.~(\ref{Rie}--\ref{lipo2}) into Eq.~\eqref{Hint}, we derive the interaction Hamiltonian describing the coupling between a plane gravitational wave and the ion crystal,
\begin{equation}
    \begin{split}
        \hat{H}_{\rm GW} \simeq \sum_i \frac{m_{\rm ion} \omega^2
        e^{i\omega t}}{8 \sqrt{2}} \Big[ &h^{(+)} \, e^{i \phi^{(+)}}
        \left( x_i^2 \cos^2 \theta + z_i^2 \sin^2 \theta - y_i^2 -
        2 x_i z_i \sin \theta \cos \theta \right) \\
        &+ 2 h^{(\times)} \, e^{i \phi^{(\times)}} \left( x_i y_i
        \cos \theta - y_i z_i \sin \theta \right) \Big] + {\rm h.c.} ,
    \end{split}
    \label{INT}
\end{equation}
where the index $i$ labels each ion in the crystal.

The collective excitation along the $z$-axis appears as Eq.~\eqref{eq:zmode},
which can be approximated by drumhead modes with the Neumann condition when a sufficient number of ions
is contained, as discussed in Appx.~\ref{sec:drumhead}.
Through the interaction~\eqref{INT}, gravitational waves can excite the phonon modes resonantly.
We focus on the linear terms in $z_i$ in Eq.~\eqref{INT}.
As the center of the ion crystal is fixed in the trap,
the summation over $x_i$ and $y_i$ in Eq.~\eqref{INT} vanishes for
parity-even phonon modes.
This is because of the quadratic nature of gravitational waves.
On the other hand, the parity-odd modes can interact with gravitational waves.

The motion of ions in the $xy$-plane is given by
\begin{equation}
    x_i  = \rho_i \cos(\omega_r t + \phi_i),
    \quad y_i = \rho_i \sin(\omega_r t + \phi_i),
    \label{xy} 
\end{equation}
where $\omega_r$ is the angular frequency of the rotation.
We further employ the continuum approximation, which is valid when a sufficiently larger number of ions are contained 
in the wavelengths of the phonon modes.
The sum of ions is evaluated as $\sum_i
\simeq (N / \pi R^2) \int r \, {\rm d}r \, {\rm d}\phi$, where the crystal
radius \( R \) scales as \( \propto \sqrt{N} \), with the lattice constant
assumed to be fixed.

From Eqs.~\eqref{INT}, \eqref{xy}, and \eqref{z}, we find that only two modes can be resonantly excited 
by gravitational waves, and other parity-odd modes are decoupled.
The interaction Hamiltonian is given by
\begin{equation}
    \hat{H}_{\rm int} = \alpha_2 \, \hat{a}_2 + \alpha_3 \, i \hat{a}_3
    + {\rm h.c.}
\end{equation}
with the amplitudes
\begin{align}
    \alpha_2 &= \frac{\sqrt{N m_{\rm ion}} (\omega_2 \pm \omega_r)^2
    R}{8 \sqrt{2} \zeta_{11} \sqrt{(\zeta_{11}^2 - 1) \, \omega_2}} \,
    \sin \theta \left[ - h^{(+)} e^{i \phi^{(+)}} \cos
    \theta \mp i h^{(\times)} e^{i \phi^{(\times)}} \right] , \\
    \alpha_3 &= \frac{\sqrt{N m_{\rm ion}} (\omega_2
    \pm \omega_r)^2 R}{8 \sqrt{2} \zeta_{11} \sqrt{(\zeta_{11}^2 - 1)
    \, \omega_2}} \, \sin \theta \left[ \pm h^{(+)} e^{i \phi^{(+)}}
    \cos \theta + i h^{(\times)} e^{i \phi^{(\times)}} \right] ,
    \label{oddd}
\end{align}
under the rotating wave approximation at the resonance $\omega = \omega_2 \pm \omega_r$.
As discussed in Appx.~\ref{sec:odf}, when the ODF frequency is specifically chosen to be
$\omega_{\rm ODF} = \omega_2 - \omega_r$, 
the interaction Hamiltonian of gravitational waves at the angular
frequency $\omega = \omega_{\rm ODF}$ is obtained
as Eq.~\eqref{alpha} with the amplitude
\begin{equation}
    \alpha = - \frac{\sqrt{N m_{\rm ion}} (\omega_2 - \omega_r)^2 R}{8
    \zeta_{11} \sqrt{(\zeta_{11}^2 - 1) \, \omega_2}} \, h_0 \, \sin
    \theta \big( \cos \theta - i \big) e^{i \phi} ,
\end{equation}
where the gravitational wave is assumed to be unpolarized, i.e.,
$\phi^{(+)} = \phi^{(\times)} \equiv \phi$ and $h^{(+)} = h^{(\times)}
\equiv h_0$.

The measurement is repeated many times, and we obtain the average of the signal parameter
$\eta = \vert {\rm Im} \, \alpha \vert$ with respect to $\theta$ and
$\phi$, that is,
\begin{equation}
    \bar{\eta} = \frac{2 \sqrt{2}}{3 \pi} \, K(1/\sqrt{2}) \times
    \frac{\sqrt{N m_{\rm ion}} (\omega_2 - \omega_r)^2 R}{8
    \zeta_{11} \sqrt{(\zeta_{11}^2 - 1) \, \omega_2}} h_0 ,
    \label{eq:etaodd}
\end{equation}
where $K (x)$ is the first complete elliptic integral, and the
numerical prefactor is around 0.556.
Here we assume that $\theta$ and $\phi$ are randomly distributed for
each gravitational wave which interacts within each measurement for
simplicity.

Using Eq.~\eqref{eq:etaodd} and Eq.~\eqref{deleta}, one can estimate the sensitivity to the amplitude of the 
gravitational wave corresponding to ${\rm SNR} = 1$ (68\% C.L.) for one-year observation as 
\begin{equation}
    h_0(f) = 4.9 \times 10^{-15} \times
    \left( \frac{N}{150} \right)^{-1/2}
    \left( \frac{m_{\rm ion}}{8.3~{\rm GeV}} \right)^{-1/2} 
    \left( \frac{R}{0.1~{\rm mm}} \right)^{-1} 
    \left( \frac{f}{1.6~{\rm MHz}} \right)^{-3/2} .
\end{equation}
Here we adopt the experimental parameters $g = 2\pi\times 3.9$~kHz,
$\sigma = 2 \pi \times 1$~Hz, and $\bar{n} = 0.3$, $R = 0.1~{\rm mm}$, following
Refs.~\cite{Gilmore:2021qqo,jordan2019near}.
The duration of a single measurement and the ODF interaction time are chosen to 
optimize the sensitivity and taken as
$T = 0.06$~s and $\tau = 0.2$~ms for the above parameters.

\begin{figure}[t!]
    \centering    
    \includegraphics[width=12cm]{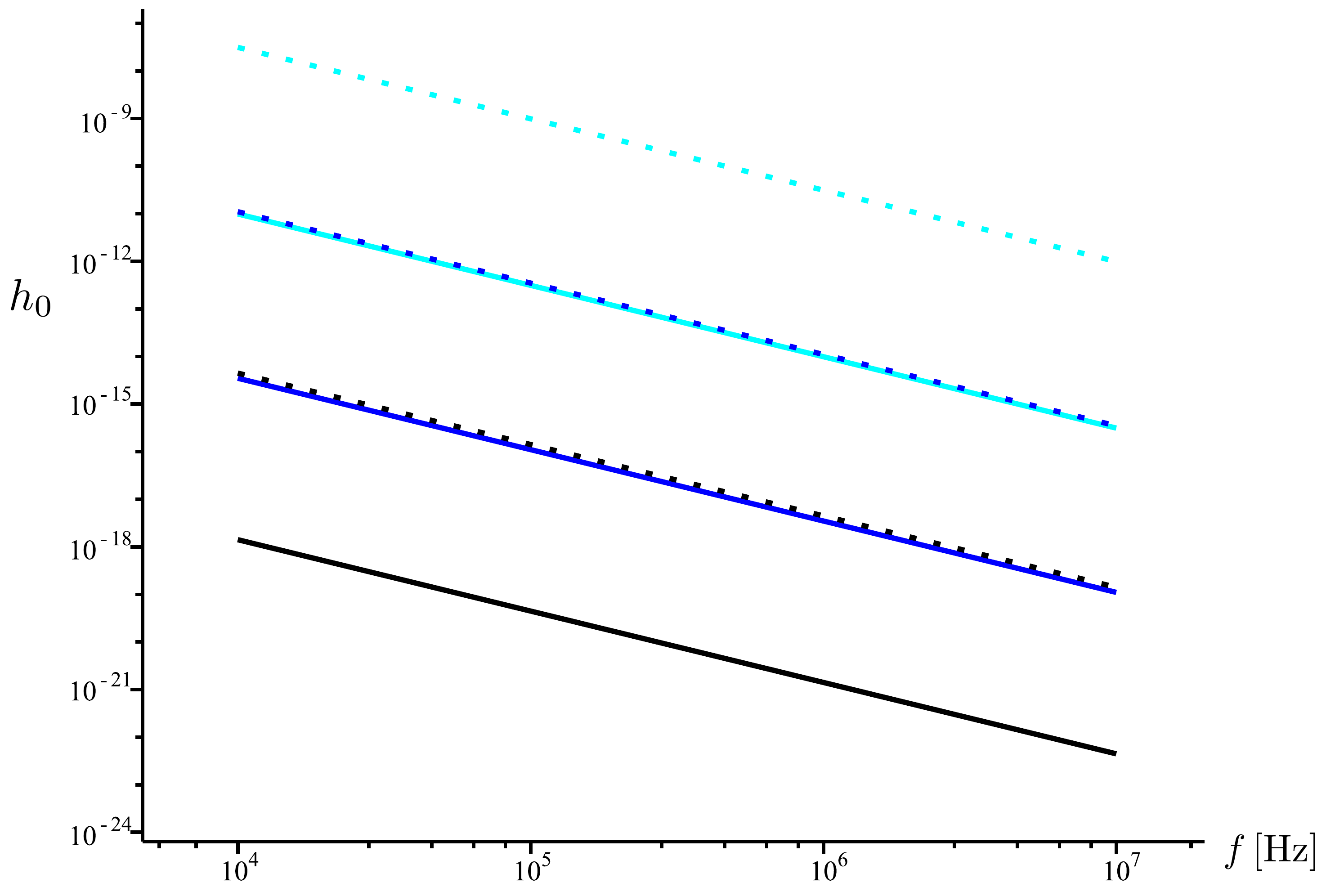}  
    \caption{
    Sensitivities to the amplitude of gravitational waves, $h_0$,
    as a function of the frequency $f$ are shown.
    Cyan, blue, and black colored lines correspond to ion numbers of $N=150$, $10^5$, $10^8$, respectively. 
    The solid curves represent the achievable sensitivity with one-year observation for each frequency bin, 
    while the dotted curves show the sensitivities obtained by scanning the entire frequency range during the total observation time of one year.
    The single measurement time $T$ and the ODF duration $\tau$ are chosen to optimize the sensitivities as
    $(T, \tau) = (0.04~{\rm s}, 0.3~{\rm ms})$, $(0.08~{\rm s}, 1~{\rm ms})$,
    and $(0.1~{\rm s}, 3~{\rm ms})$ for 
    $N=150$, $10^5$, and $10^8$, respectively.
    }
    \label{figh0odd}
\end{figure}

\begin{figure}[t!]
    \centering    
    \includegraphics[width=12cm]{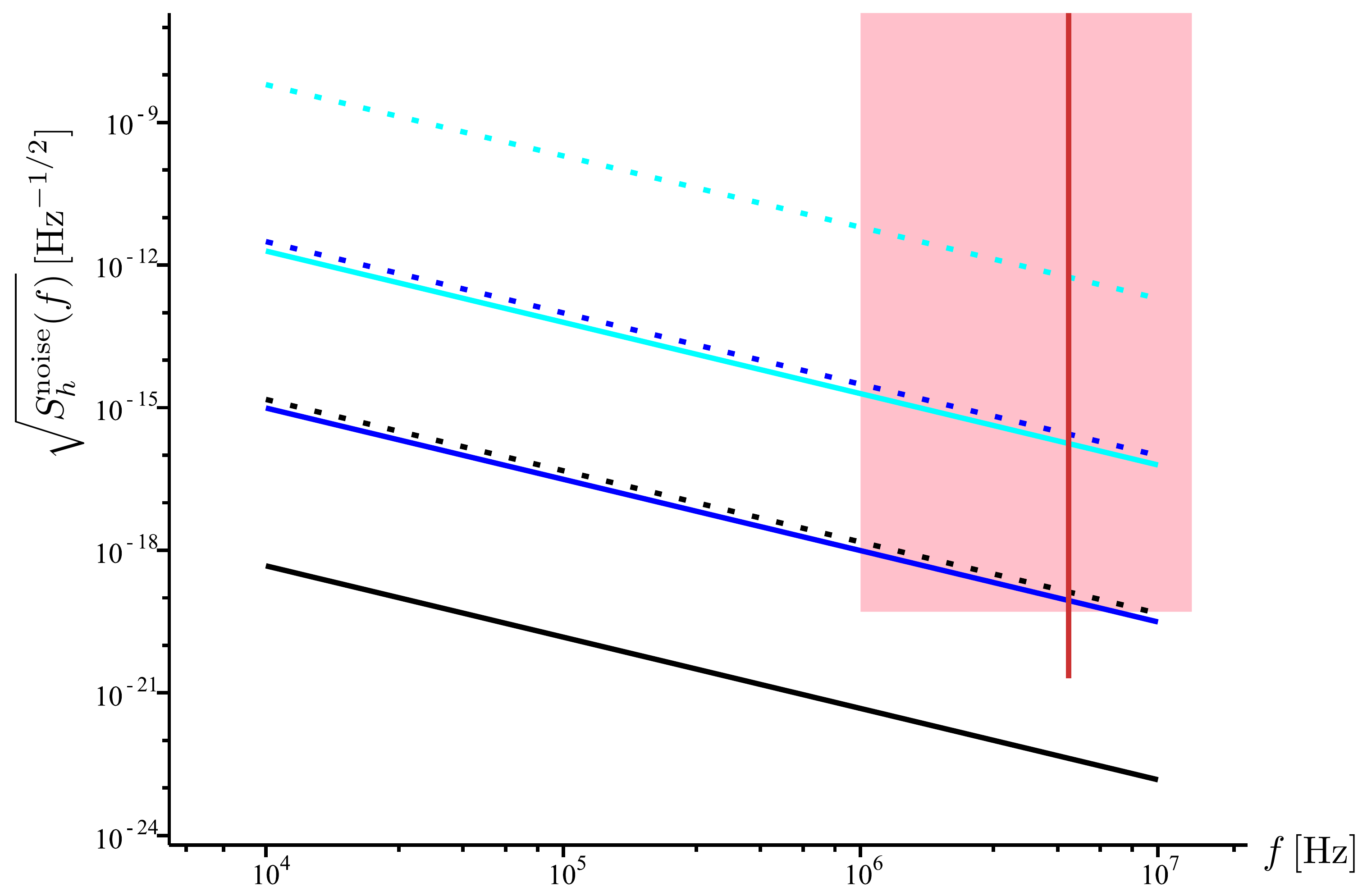}  
    \caption{
    Sensitivities to the noise-equivalent spectral density of gravitational waves, 
    $S_h^{\rm noise}(f)$, as a function of the frequency $f$ are shown.
    The experimental parameters are the same as Fig.~\ref{figh0odd}.
    The pink band and the red line express the existing experiments,
    the Fermilab Holometer~\cite{Holometer:2016qoh} and the Bulk
    Acoustic Wave (BAW) experiments~\cite{Goryachev:2021zzn}.
    }
    \label{figShodd}
\end{figure}

In Fig.~\ref{figh0odd}, we show the sensitivities to the amplitude of gravitational waves, 
\( h_0(f) \), for the cases of \( N = 150\), \( 10^5 \), and \( 10^8 \).
The solid curves represent the achievable sensitivity with one-year observation per frequency bin, 
while the dotted curves show the sensitivities obtained by scanning the entire frequency range for a year. 
The target frequency can be tuned within
$f \in [10~\mathrm{kHz},\, 10~\mathrm{MHz}]$
by adjusting \( \omega_z \)~\cite{Gilmore:2021qqo}.
The frequency range is related to the stability of the ion trap system.
Once we assume the use of realistic voltages and magnetic fields, the
frequency of the axial oscillation is limited in such range.
For each pair of sensitivity curves, the single measurement time $T$
and the ODF duration $\tau$ are chosen as
$(T, \tau) = (0.04~{\rm s}, 0.3~{\rm ms})$, $(0.08~{\rm s}, 1~{\rm ms})$,
and $(0.1~{\rm s}, 3~{\rm ms})$ for 
$N = 150$, $10^5$, and $10^8$.
We scaled the crystal size $R$ as $R \propto \sqrt N$.

It may be more convenient to use the power spectral density of gravitational waves, which would be related to 
the amplitude of a continuous gravitational wave as $S_h(f) = T h_0^2$~\cite{Aggarwal:2025noe}.
The noise-equivalent spectral density is defined as~\cite{Aggarwal:2025noe}
\begin{equation}
   \sqrt{S_h^{{\rm noise}}(f)} = \frac{\sqrt{S_h(f)}}{{\rm SNR}} \, .
\end{equation}
If one considers the odd mode of the ion crystal with \( N = 10^8 \),
the noise-equivalent spectral density can be estimated as
\begin{equation}
    \sqrt{S_h^{{\rm noise}}(f)} = 2.3 \times 10^{-22}~{\rm Hz}^{-1/2}
    \times \left( \frac{N}{10^8} \right)^{-1/2}
    \left( \frac{m_{\rm ion}}{8.3~{\rm GeV}} \right)^{-1/2} 
    \left( \frac{R}{80~{\rm mm}} \right)^{-1} 
    \left( \frac{f}{1.6~{\rm MHz}} \right)^{-3/2}.
\end{equation}
with $(T, \tau) = (0.1~{\rm s}, 2~{\rm ms})$.
In Fig.~\ref{figShodd}, the sensitivities to the noise-equivalent spectral density 
\( S_h^{\mathrm{noise}}(f) \) for ion numbers of \( N = 150 \), \( 10^5 \), and \( 10^8 \)
are shown in comparison with other experiments, 
such as the Fermilab Holometer~\cite{Holometer:2016qoh} and the 
Bulk Acoustic Wave (BAW) experiments~\cite{Goryachev:2021zzn}.
The same experimental parameters as in Fig.~\ref{figh0odd} are used in Fig.~\ref{figShodd}.

Our method is based on resonant excitations, and the sensitivities shown as solid lines in 
Figs.~\ref{figh0odd} and \ref{figShodd} represent the achievable reach at a given frequency, 
with the linewidth determined by the quality factor, which is 
\( Q \sim 10^4 \) in our setup.
We adopt this convention and show the sensitivities with solid lines, 
following Ref.~\cite{Aggarwal:2025noe}.
If one considers scanning a certain frequency range, multiple detectors or a longer total observation time
would be required to maintain the sensitivity.
As an illustration, in the case of a single detector with the total observation time fixed at one year, the sensitivities obtained by scanning the entire frequency range, $f \in [10~\mathrm{kHz},\, 10~\mathrm{MHz}]$, are shown by the dotted curves 
in Figs.~\ref{figh0odd} and \ref{figShodd}.


\section{Conclusions}
\label{sec:conclusion}

We study a detection method for high-frequency gravitational waves by
using two-dimensional ion crystals, where ions are trapped by a
magnetic field and an electric potential, and it exhibits collective
oscillations known as drumhead modes.
It is shown that parity-odd modes can be excited by gravitational waves
due to their quadrupole nature.
This feature provides a unique signature of gravitational waves, useful for discriminating them from other sources.

We investigate a detection protocol for such excitations using the ODF, which allows the excitation to be readout through the rotation of the total spin. 
As a squeezed spin state is generated through this protocol, 
gravitational wave detection beyond the standard quantum limit becomes possible.
Furthermore, choosing observational parameters, such as the total observation time
and the duration of the ODF operation, can optimize the
signal-to-noise ratio.
The sensitivity using the first odd mode could be comparable to or better than other experiments.
Further improvement of the sensitivity can be expected by reducing the thermal noise and instability of the ODF laser.

We consider planar gravitational waves in the high-frequency range of 
$10$~kHz--$10$~MHz.
Such high-frequency gravitational waves can be generated, for example, 
from binaries or the superradiance of light primordial
black holes~\cite{Franciolini:2022htd}.
On the other hand, 
stochastic gravitational waves in the high-frequency regime could also be produced in the early universe by
processes such as a first-order phase transition around
$10^7$--$10^{10}$~GeV and the decay of topological 
solitons~\cite{Guo:2025cza}.
Since stochastic gravitational waves can be described by a superposition of planar gravitational waves,
the sensitivity estimates obtained in this paper would be applicable to them as well, 
at least as an order-of-magnitude estimate.
Moreover, 
continuous and stochastic gravitational waves can, in principle, be distinguished by observing 
their directional dependence and/or frequency spectra.

Although we focus on two-dimensional ion crystals in this paper,
it would be interesting to extend the discussion to three-dimensional crystals.
We also discuss the possibility of using the center-of-mass mode for gravitational wave detection, assuming that the ion crystal undergoes orbital motion, in Appx.~\ref{sec:com}. The sensitivity for the case of graviton-photon conversion is also shown there. 
Investigating these possibilities from both theoretical and experimental
perspectives is interesting, and we leave these issues for future work.


\section*{Acknowledgment}

This work was in part supported by World Premier International
Research Center Initiative (WPI), MEXT, Japan, and JSPS KAKENHI Grant
Numbers JP22K14034 (A.I.), JP22K21350 (R.K.), JP23KJ2173 (W.N.) and JP24KJ1157 (R.T.).


\appendix
\numberwithin{equation}{section}
\setcounter{equation}{0}


\section{Drumhead modes}
\label{sec:drumhead}

From Eq.~\eqref{pote}, the equation of motion along the $z$-axis for the
$i$th ion in the Penning trap potential is
\begin{align}
    m_{\rm ion} \ddot{z}(\bm{\rho}_i,t) = - m_{\rm ion} \omega_z^2 \,
    z(\bm{\rho}_i,t) + \alpha_{\rm EM} \sum_j \frac{z(\bm{\rho}_i,t)
    - z(\bm{\rho}_i + \bm{\delta}_{ij},t)}{|\bm{\delta}_{ij}|^3} , 
\end{align}
where $\bm{\rho}_i$ denotes the position of the $i$th ion in the $xy$-plane, and $\bm{\delta}_{ij}$
represents the vector pointing from the $i$th ion to the $j$th ion.
For a sufficiently large number of ions, one can take the continuum limit, and the equation is rewritten as
\begin{align}
    \ddot{z}(\bm{\rho},t) \simeq -\omega_z^2 \, z(\bm{\rho},t) - v^2 \nabla^2 z(\bm{\rho},t) , \label{zeq}
\end{align}
where we define $v^2 = (\alpha_{\rm EM} / 2 m_{\rm ion}) \sum_j \vert
\bm{\delta}_{ij} \vert^{-1}$, assumed to be independent of $i$.
In the derivation, $\sum_j \bm{\delta}_{ij} / \vert \bm{\delta}_{ij}
\vert^3 \cdot \nabla z(\bm{\rho}_i,t) \simeq 0$ is used due to the
local symmetry of the ion crystal. 
This symmetry is violated near the edge of the crystal, but
instead ions around the rim satisfy $\nabla z(\bm{\rho},t) \simeq 0$,
as we can see later.

The solution of Eq.~\eqref{zeq} is given by $z(\bm{\rho},t) =
P (\rho) \, \Phi(\phi) \, T (t)$ with
\begin{equation}
    P (\rho) = J_m \left( \zeta_{mn} \frac{\rho}{R} \right) , \quad
    \Phi(\phi) = A \cos m \phi + B \sin m \phi , \quad
    T (t) = \sin \omega_{mn} t .
\end{equation}
Here $J_m (x)$ is the $m$th Bessel function, $\zeta_{mn}$ is the
$n$th zero of its derivative $J'_m (x)$, and $A$, $B$ are the
integration constants.
The frequency $\omega_{mn}$ is defined as
\begin{equation}
  \omega_{mn} = \sqrt{ \omega_z^2 - \frac{\zeta_{mn}^2 v^2}{R^2} } . \label{a7}
\end{equation}
We impose the Neumann boundary condition, i.e., $P' (\rho = R) = 0$,
and the initial condition $z ({\bm \rho}, t = 0) = 0$ at any
${\bm \rho}$.
For example, the frequency of the center-of-mass mode is $\omega_1 = \omega_{01} = \omega_z$, which is the highest mode.
The succeeding modes are degenerate and given by 
$\omega_2 = \omega_3 = \omega_{11}$, $\omega_4 = \omega_5 = \omega_{21}$, $\dots$.

For an ion crystal consisting of 272 $^9 \mathrm{Be}^+$ ions, 
the drumhead modes correctly reproduce the ordering of the frequencies obtained without the 
continuum approximation
up to $\omega_k \ge \omega_5$~\cite{sawyer2012spectroscopy,britton2012engineered}. 
In order to reproduce the correct ordering of the lower-frequency modes, a larger number of ions is required to validate the continuum approximation, where a sufficient number of ions are contained
within the wavelength of each mode.

We can quantize $z (\bm{\rho})$ as
\begin{align}     \label{z}
    \hat{z}(\bm{\rho}) &= \frac{1}{\sqrt{2 m_{\rm ion}
    \omega_{01}}} \frac{1}{\sqrt{N}} \left( \hat{a}_{01} + \hat{a}_{01}^\dagger
    \right) \nonumber \\ 
    &\quad + \sum_{m, n} \frac{1}{\sqrt{2 m_{\rm ion} \omega_{mn}}} \frac{1}{\sqrt{N}}
    \sqrt{\frac{2}{\zeta_{mn}^2 - m^2}} \frac{\zeta_{mn}}{J_m (\zeta_{mn})}
    J_m \left( \frac{\zeta_{mn} \rho}{R} \right) \\
    &\qquad \times \bigg[ \cos (m \phi) \left( \hat{a}_{mn} +
    \hat{a}_{mn}^\dagger \right) + \sin (m \phi) \left( \hat{a}_{-mn} +
    \hat{a}_{-mn}^\dagger \right)  \bigg] \nonumber ,
\end{align}
where the sums of $m \ge 0$ and $n$ are taken from the second highest frequency
mode $(1, 1)$ to the lower-frequency modes. 
Comparing this expression with Eq.~\eqref{eq:zmode}, one can see how
the phonon eigenvectors can be written in the continuum limit for each mode.
For example, for the center-of-mass mode $(k = 1)$ corresponding to $(m, n) = (0, 1)$,
the eigenvector is given by $b_{i1} = b_{i(01)} = 1/\sqrt{N}$.
For the $k = 2, 3$ modes, expressed respectively as $\hat{a}_{11}$ and
$\hat{a}_{-11}$, we find
\begin{align}
    b_{i2} = b_{i(11)} &= \frac{1}{\sqrt{N}}
    \sqrt{\frac{2}{\zeta_{11}^2 - 1}} \frac{\zeta_{11}}{J_1 (\zeta_{11})}
    J_1 \left( \frac{\zeta_{11} \rho_i}{R} \right) \cos \phi_i , \\
    b_{i3} = b_{i(-11)} &= \frac{1}{\sqrt{N}}
    \sqrt{\frac{2}{\zeta_{11}^2 - 1}} \frac{\zeta_{11}}{J_1 (\zeta_{11})}
    J_1 \left( \frac{\zeta_{11} \rho_i}{R} \right) \sin \phi_i .
\end{align}
Three vectors ${\bm b}_1$, ${\bm b}_2$, and ${\bm b}_3$ are orthogonal to each
other and normalized in the continuum limit.


\section{Inhomogeneous optical dipole force}
\label{sec:odf}

In this appendix, we derive the Hamiltonian of the inhomogeneous ODF, Eq.~\eqref{eq:odf2}. We start from the AC stark
shift~\cite{britton2012engineered,Polloreno:2022nxl}
\begin{align}
    H_{\rm ODF} &= \frac{U}{2} \sum_i \sin \big(
    \delta k \, z_i - \omega_{\rm ODF} t - \phi_{\rm ODF} -
    f_{\rm lab} (\bm{\rho}_i, t) \big) \sigma_i^z \\
    &\supset \frac{U \delta k}{2} \sum_i \cos \big(
    \omega_{\rm ODF} t + \phi_{\rm ODF} + f_{\rm lab} (\bm{\rho}_i, t)
    \big) z_i \sigma_i^z ,
\end{align}
where $U$ is the strength of the shift, and $\delta k$ and
$\phi_{\rm ODF}$ are the $z$-component of the wavevector and the initial
phase of ODF, respectively.

The local phase shift in the laboratory frame, $f_{\rm lab}(\bm{\rho}_i,t) =
f(\rho_i) \cos \big( \omega_r t + \phi_i \big)$, is constructed
by deformable mirrors~\cite{Polloreno:2022nxl}.
Hereafter, $\omega_{\rm ODF} = \omega_n - \omega_r + \delta$ is chosen, where $\delta$ is the detuning of the ODF lasers assumed to be constant during a single measurement
and normally distributed with $\langle \delta \rangle = 0$ and $\langle
\delta^2 \rangle = \sigma^2$~\cite{Gilmore:2021qqo}.
In the interaction picture of $\hat{H}_0 = \sum_k \omega_k \hat{a}_k^\dagger \hat{a}_k + \omega_0 \hat{J}_z$, where $\omega_0$ is the energy gap of the spin states, the ODF term becomes
\begin{align}
    \hat{H}_{{\rm ODF},{\rm I}} &\simeq \sum_i \sum_k
    \frac{U \delta k \, b_{ik}}{4 \sqrt{2 m_{\rm ion} \omega_k}} \left(
    \hat{a}_k e^{-i\omega_k t} + \hat{a}_k^\dagger e^{i\omega_k t} \right)
    \hat{\sigma}_i^z \nonumber \\
    &\quad \times \left[ e^{ i(\omega_{\rm ODF} t + \phi_{\rm ODF})} \sum_{\ell=-\infty}^\infty i^\ell J_\ell(f(\rho_i)) \, e^{i\ell (\omega_r t + \phi_i)} + {\rm c.c.} \right] \\
    &\simeq \sum_i \sum_{\{ k \, \vert \, \omega_k = \omega_n \}} \frac{U \delta k \, b_{ik}}{4 \sqrt{2 m_{\rm ion} \omega_n}} J_1 \big( f(\rho_i) \big) \left[ \hat{a}_k e^{i (\delta t + \phi_i)} + {\rm h.c.} \right] \hat{\sigma}_i^z ,
    \label{eq:odf_I}
\end{align}
where the Jacobi--Anger expansion $e^{iz \cos\phi} = \sum_{\ell=-\infty}^\infty i^\ell J_\ell(z) e^{i\ell \phi}$ and the rotating wave approximation is employed.
We set the ODF phase $\phi_{\rm ODF} = - \pi / 2$.
Henceforth the detuning $\delta$ is inserted into the free Hamiltonian to
cancel out the time dependence of the ODF~\eqref{eq:odf_I}.

For $n = 2$, we set $f(\rho_i) = \zeta_{11} \rho_i / R$.
Defining another interaction picture by $\hat{H}_0 + \delta \sum_{k = 2}^3 \hat{a}_k^\dagger \hat{a}_k$, the inhomogeneous ODF is written as 
\begin{equation}
    \begin{split}
        \hat{H}_{{\rm ODF},{\rm I}} = &- \delta \sum_{k = 2}^3
        \hat{a}_k^\dagger
        \hat{a}_k + \frac{g_2}{\sqrt{N}} (\hat{a}_2+\hat{a}_2^\dagger)
        \hat{J}^z_2 + \frac{i g_2}{\sqrt{N}} (\hat{a}_3 - \hat{a}_3^\dagger)
        \hat{J}^z_3 \\
        &+ \frac{i g_2}{\sqrt{N}} (\hat{a}_2 - \hat{a}_2^\dagger) \hat{J}_{23}^z +
        \frac{g_2}{\sqrt{N}} (\hat{a}_3 + \hat{a}_3^\dagger) \hat{J}_{23}^z ,
    \end{split}
    \label{eq:odf_int2}
\end{equation}
with
\begin{equation}
    g_2 = \sqrt{\frac{\zeta_{11}^2 - 1}{2}} \frac{J_1 (\zeta_{11})}{\zeta_{11}}
    \frac{U \, \delta k}{2 \sqrt{2 m_{\rm ion} \omega_2}}, \quad
    \hat{J}_k^z = \sum_i N b_{ik}^2 \frac{\hat{\sigma}_i^z}{2}, \quad
    \hat{J}_{23}^z = \sum_i N b_{i2} b_{i3} \frac{\hat{\sigma}_i^z}{2} .
\end{equation}
We extract the total component $\hat{J}_z$ from $\hat{J}_k^z$ as
$\hat{J}_k^z = \hat{J}_z + \sum_i (N b_{ik}^2 - 1) \, \hat{\sigma}_i^z/2$.
The second term does not contribute to the expectation value of the
total spin after the protocol shown in Sec.~\ref{sec:odd_protocol}.
The terms containing $\hat{J}_{23}^z$ have sub-leading contributions of $1/N$ depending on noise effects, and therefore
they are ignored in the main text.


\section{Noise effect on sensitivity}
\label{sec:noise}

In this appendix, we produce Eq.~\eqref{eq:odd_eta0}, the sensitivity
$\delta \eta$ with noises.
We need two expectation values $\langle \hat{J}_z \rangle$
and $\langle \hat{J}_z^2 \rangle$, starting from the state $\Ket{{\rm init}}$ to the
state $\Ket{{\rm fin}}$, as shown in Sec.~\ref{sec:odd_protocol}.
This is equivalent to the calculation of two values $\langle \hat{J}_y (T) \rangle$
and $\langle \hat{J}_y^2 (T) \rangle$ in the state $\Ket{\psi_0} = \Ket{N/2}_x$ in
the Heisenberg picture, except for the state preparation and measurement error.
To include thermal noise in the initial state, we start with the density
matrix $\hat{\rho}_0 = \Ket{\psi_0} \Bra{\psi_0} \otimes \hat{\rho}_{\bar{n}}$,
where $\hat{\rho}_{\bar{n}} = \sum_{n = 0}^\infty p_{n} \Ket{n} \Bra{n}$ is the thermal
state density of $\hat{a}^{(\dagger)}$ satisfying $\sum_{n} n p_{n} = \bar{n}$. 


\subsection{Time evolution of spin operators}

The time-dependent Hamiltonian is described as
\begin{equation}
    \hat{H} (t) = - \delta \hat{a}^\dagger \hat{a} +
    \frac{g (t)}{\sqrt{N}} \big( \hat{a}^\dagger + \hat{a} \big)
    \hat{J}_z + \left( \alpha (t) \, e^{-i \delta t} \, \hat{a} +
    \alpha^* (t) \, e^{i \delta t} \, \hat{a}^\dagger \right) ,
    \label{eq:noise_H}
\end{equation}
with a real function $g (t)$ signifying that there is switching of the ODF
and $\alpha (t)$ denoting the contribution of gravitational waves including frequency detuning in the interaction picture~\cite{wall2017boson}.
The total spin operator $\hat{J}_z$ is conserved.
There is an assumption that the noise model is in good approximation
for the single measurement time~\cite{Gilmore:2021qqo}.

The equations of motion for the phonon annihilation operator and the spin
ladder operator in the Heisenberg picture are given by
\begin{align}
    i \frac{{\rm d}}{{\rm d}t} \hat{a} (t) &= \left[ \hat{a} (t),
    \hat{H} (t) \right] = -\delta \hat{a} (t) + \frac{g (t)}{\sqrt{N}}
    \, \hat{J}_z + \alpha^* (t) \, e^{i \delta t} , \\
    i \frac{{\rm d}}{{\rm d}t} \hat{J}_\pm (t) &= \left[ \hat{J}_\pm
    (t), \hat{H} (t) \right] = \mp \frac{g (t)}{\sqrt{N}} \left(
    \hat{a} (t) + \hat{a}^\dagger (t) \right) \hat{J}_\pm (t) ,
\end{align}
which are solved as
\begin{align}
    \hat{a} (t) &= \hat{a} \, e^{i \delta t} - i
    \frac{G^*_\delta (t)}{\sqrt{N}} e^{i \delta t} \hat{J}_z
    - i A^* (t) \, e^{i \delta t}, \\ 
    \hat{J}_\pm (t) &= \exp \left[ \pm \frac{i}{\sqrt{N}}
    \left( G_\delta (t) \, \hat{a} + G^*_\delta (t) \, \hat{a}^\dagger -
    \frac{2 H_\delta (t)}{\sqrt{N}} \left( \hat{J}_z \mp \frac{1}{2}
    \right) + 2 B_\delta (t) \right) \right] \hat{J}_\pm ,
\end{align}
with
\begin{equation}
    \begin{split}
        A (t) &= \int_0^t {\rm d}t' \, \alpha (t') , \quad
        B_\delta (t) = - \, {\rm Im} \int_0^t {\rm d}t' \, A (t')
        \frac{{\rm d} G^*_\delta (t')}{{\rm d}t'} , \\
        G_\delta (t) &= \int_0^t {\rm d}t' \, g (t') e^{i \delta t'} ,
        \quad H_\delta (t) = {\rm Im} \int_0^t {\rm d}t' \,
        G_\delta (t') \frac{{\rm d} G^*_\delta (t')}{{\rm d}t'} .
    \end{split}
\end{equation}
The effect of gravitational waves is encoded in the functions $A (t)$ and
$B_\delta (t)$.

The expectation values of $\hat{J}_\pm (t)$ and $\hat{J}_\pm^2 (t)$ are given by
\begin{equation}
    \langle \hat{J}_\pm (t) \rangle = \frac{N}{2}
    \exp \left[- \frac{\vert G_\delta (t) \vert^2}{N} \left( \bar{n} +
    \frac{1}{2} \right) \pm \frac{2 i}{\sqrt{N}} B_\delta (t) \right]
    \cos^{N-1} \frac{H_\delta (t)}{N} ,
    \label{eq:ev_jpm}
\end{equation}
and
\begin{equation}
    \langle \hat{J}^2_\pm (t) \rangle = \frac{N (N-1)}{4}
    \exp \left[- \frac{4 \, \vert G_\delta (t) \vert^2}{N} \left( \bar{n} +
    \frac{1}{2} \right) \pm \frac{4 i}{\sqrt{N}} B_\delta (t) \right]
    \cos^{N-2} \frac{2 H_\delta (t)}{N} ,
    \label{eq:ev_jpmSq}
\end{equation}
respectively.


\subsection{Spin dissipation}

The spin dissipation effect is described as a master equation
\begin{equation}
    \frac{{\rm d}}{{\rm d} t} \hat{\rho} (t) = i \left[ \hat{\rho}(t),
    \hat{H}(t) \right] +\frac{\Gamma}{4} \sum_i \Big( \hat{\sigma}^z_i
    \hat{\rho}(t) \hat{\sigma}^z_i - \hat{\rho}(t) \Big) ,
\end{equation}
where $\Gamma$ is the dissipation rate of the spins by the ODF photon scattering. The dissipation operator commutes with the Hamiltonian, and thus the equation is solved separately. To show the dissipation effect for the expectation values, $\hat{H}(t)=0$ is taken for simplicity and this case is distinguished by taking the subscript I. The equation becomes
\begin{align}
    \frac{{\rm d}}{{\rm d} t} \hat{\rho_{\rm I}} (t) =
    \frac{\Gamma}{4} \sum_i \Big( \hat{\sigma}^z_i \,
    \hat{\rho_{\rm I}}(t) \, \hat{\sigma}^z_i - \hat{\rho_{\rm I}}(t)
    \Big) .
\end{align}

Then, the expectation value of the spin ladder operator $\hat{J}_{\pm} = \hat{J}_x \pm i \hat{J}_y = \sum_i \hat{\sigma}^\pm_i$ for the transferred density matrix, $\langle \hat{J}_\pm \rangle_{\rm I} (t)$, decays as
\begin{equation}
    \frac{{\rm d}}{{\rm d}t} \langle \hat{J}_\pm \rangle_{\rm I} (t) =
    {\rm Tr} \left[ \hat{J}_\pm \, \frac{{\rm d}}{{\rm d}t}
    \hat{\rho_{\rm I}} (t) \right] = - \frac{\Gamma}{2} \langle
    \hat{J}_\pm \rangle_{\rm I} (t) .
\end{equation}
We obtain
\begin{equation}
    \langle \hat{J}_\pm^n \rangle_{\rm I} (t) = e^{-n \Gamma t / 2}
    \langle \hat{J}_\pm^n \rangle_{\rm I} (0)
    \label{eq:dis}
\end{equation}
for $n = 1$, 2.
The dissipation effect during a single measurement is $e^{-n \Gamma
\tau}$ because the duration of the ODF operation is $2\tau$.
Similarly,
\begin{align}
    \left\langle \hat{J}_+ \hat{J}_- + \hat{J}_- \hat{J}_+ \right\rangle_{\rm I} (t) = N + \left( \left\langle \hat{J}_+ \hat{J}_- + \hat{J}_- \hat{J}_+ \right\rangle_{\rm I} (0) - N \right) e^{-\Gamma t}
\end{align}
is obtained.
Again, the dissipation factor is given by $e^{-2 \Gamma \tau}$ for the
measurement.
If $\Gamma = 0$, this operator is conserved as $\langle \hat{J}_+ (t)
\hat{J}_- (t) + \hat{J}_- (t) \hat{J}_+ (t) \rangle = 2 \langle
\hat{\bm{J}}^2 - \hat{J}_z^2 \rangle = N (N+1) / 2$, and therefore
we find
\begin{equation}
    \left\langle \hat{J}_+ \hat{J}_- + \hat{J}_- \hat{J}_+ \right\rangle
    (T) = N + \frac{N (N-1)}{2} e^{-2 \Gamma \tau} .
    \label{eq:ev_jjjj}
\end{equation}


\subsection{Sensitivity}

Combining Eq.~\eqref{eq:ev_jpm} and the dissipation error~\eqref{eq:dis},
we obtain
\begin{align}
    \langle \hat{J}_y (T) \rangle &= \frac{N}{2} \exp
    \left[- \Gamma \tau - \frac{\vert G_\delta (T) \vert^2}{N} \left(
    \bar{n} + \frac{1}{2} \right) \right] \sin \frac{2 B_\delta
    (T)}{\sqrt{N}} \cos^{N-1} \frac{H_\delta (T)}{N} \\
    &= \sqrt{N} B_\delta (T) \exp \left[-\Gamma \tau -
    \frac{\vert G_\delta (T) \vert^2}{N} \left( \bar{n} + \frac{1}{2}
    \right) \right] \cos^{N-1} \frac{H_\delta (T)}{N} , \label{16}
\end{align}
up to the linear order of $\alpha$.
As expected, $\langle \hat{J}_y(t) \rangle$ vanishes at $\alpha = 0$.
In addition, from Eqs.~\eqref{eq:ev_jpmSq} and \eqref{eq:ev_jjjj}, we get
\begin{align}
    \langle \hat{J}_{y}^2 (T) \rangle &= \frac{N}{4} +
    \frac{N (N-1)}{8} e^{-2 \Gamma \tau} \nonumber \\
    &\quad \times \left( 1 - \exp \left[- \frac{4 \, \vert G_\delta (T)
    \vert^2}{N} \left( \bar{n} + \frac{1}{2} \right) \right] \cos \frac{4
    B_\delta (T)}{\sqrt{N}} \cos^{N-2} \frac{2 H_\delta (T)}{N} \right) \\
    &= \frac{N}{4} + \frac{N (N-1)}{8} e^{-2 \Gamma \tau} \nonumber \\
    &\quad \times \left( 1 - \exp \left[- \frac{4 \, \vert G_\delta (T)
    \vert^2}{N} \left( \bar{n} + \frac{1}{2} \right) \right] \cos^{N-2}
    \frac{2 H_\delta (T)}{N} \right) , \label{18}
\end{align}
by ignoring the ${\cal O} (\alpha^2)$ terms.

Next, we proceed to a calculation that reflects the experiment conditions. 
We perform the ODF $g (t) = g$ from $t = 0$ to $t = \tau$, and its inverse
$g (t) = - g$ from $t = T - \tau$ to $t = T$.
Thus,
\begin{equation}
    G_\delta (t) = \begin{dcases}
        - \frac{ig}{\delta} \left( e^{i \delta t} - 1 \right)
        & (0 \leq t < \tau) \\
        - \frac{ig}{\delta} \left( e^{i \delta \tau} - 1 \right)
        & (\tau \leq t \leq T - \tau) \\
        - \frac{ig}{\delta} \left( e^{i \delta \tau} - 1 \right)
        + \frac{ig}{\delta} \left( e^{i \delta t} - e^{i \delta
        (T - \tau)} \right) & (T - \tau < t \leq T)
    \end{dcases}
\end{equation}
and
\begin{equation}
    \vert G_\delta (T) \vert^2 = \frac{16 g^2}{\delta^2} \sin^2
    \frac{\delta \tau}{2} \sin^2 \frac{\delta (T - \tau)}{2}
    = \delta^2 g^2 \tau^2 (T - \tau)^2 + {\cal O} (\delta^4) .
\end{equation}
Here we assume that $\delta$, or its standard deviation $\sigma$, is much
smaller than $\tau^{-1}$ and $T^{-1}$.
Moreover,
\begin{equation}
    H_\delta (T) = - \frac{2 g^2}{\delta^2} \left[ \delta \tau -\sin \delta
    \tau -2 \sin^2 \frac{\delta \tau}{2} \sin \delta (T - \tau) \right] = \delta
    g^2 \tau^2 \left( T - \frac{4}{3} \tau \right) + {\cal O} (\delta^3) .
\end{equation}

If a gravitational wave frequency, $\omega$, is detuned from $\omega_k - \omega_r$
by $\Delta = \omega - \omega_k + \omega_r$, the signal coupling is expressed as
$\alpha (t) = \alpha e^{-i \Delta t}$.
The function $A (t)$ is given by $A (t) = (i \alpha / \Delta) \left(
e^{-i \Delta t} - 1 \right)$, and therefore we obtain 
\begin{equation}
    B_\delta (T) = \frac{g}{\Delta} \left[ ({\rm Im} \, \alpha)
    \Big( C (\Delta + \delta) - C (\delta) \Big) - ({\rm Re} \, \alpha)
    \Big( S (\Delta + \delta) - S (\delta) \Big) \right],
\end{equation}
with
\begin{equation}
    S (\delta) = \frac{4}{\delta} \sin \frac{\delta (T - \tau)}{2}
    \sin \frac{\delta \tau}{2} \sin \frac{\delta T}{2} , \quad
    C (\delta) = \frac{4}{\delta} \sin \frac{\delta (T - \tau)}{2}
    \sin \frac{\delta \tau}{2} \cos \frac{\delta T}{2} .
\end{equation}
There are two cases of detuning of gravitational waves, $\Delta$:
$\Delta \ll T^{-1}$ and $\Delta \gg \sigma$.
The former includes the resonance $\Delta = 0$, and the latter is
useful to discuss the width and the quality factor.

In the near-resonance case, $\Delta \ll T^{-1}$, we can expand $B_\delta (T)$
with respect to $\Delta$ and $\delta$ as
\begin{equation}
    \begin{split}
        B_\delta (T) &= g \tau (T - \tau) \left[ ({\rm Im} \, \alpha) -
        \frac{\Delta T}{2} ({\rm Re} \, \alpha) \right] \left[ 1 -
        \frac{1}{12} \left( \Delta^2 + 3 \delta^2 \right)
        \left( 2 T^2 - \tau T + \tau^2 \right) \right] \\
        &\quad + \big( \delta \text{-linear} \big)
        + {\cal O} \left( (\Delta, \delta)^3 \right) .
    \end{split}
\end{equation}
Taking the average over detuning $\delta$, we find
\begin{equation}
    \begin{split}
        \frac{\partial}{\partial \eta} \lvert \langle \hat{J}_y (T)
        \rangle \rvert &\simeq \sqrt{N} g \tau (T - \tau) e^{-\Gamma
        \tau} \Bigg[ 1 - \frac{1}{12} \left( \Delta^2 + 3 \sigma^2
        \right) \left( 2 T^2 - \tau T + \tau^2 \right) \\
        &\qquad - \frac{\sigma^2 g^2 \tau^2 (T - \tau)^2}{N} \left(
        \bar{n} + \frac{1}{2} \right) - \frac{N-1}{2} \frac{\sigma^2
        g^4 \tau^4}{N^2} \left( T - \frac{4}{3} \tau \right)^2 \Bigg],
    \end{split}
\end{equation}
and
\begin{equation}
    {\rm var} \big( \hat{J}_y (T) \big) \simeq \frac{N}{4} +
    \frac{N-1}{4} \sigma^2 g^2 \tau^2 (T - \tau)^2 e^{-2 \Gamma \tau}
    \left[ 2 \bar{n} + 1 + \frac{N-2}{N} \frac{g^2 \tau^2 (T -
    \frac{4}{3} \tau)^2}{(T - \tau)^2} \right] .
\end{equation}
Here, we define 
\begin{equation}
    \eta = \left\vert {\rm Im} \left( \alpha e^{-i \Delta T / 2} \right)
    \right\vert \simeq \left\vert ({\rm Im} \, \alpha) - \frac{\Delta T}{2}
    ({\rm Re} \, \alpha) \right\vert .
\end{equation}
The sensitivity for a single measurement is finally given by 
\begin{align}
    (\delta \eta)^2 
    &= \frac{{\rm var} \big( \hat{J}_y (T) \big)}{\left(\frac{\partial}{\partial \eta} \lvert \langle \hat{J}_y (T)
        \rangle \rvert \right)^2} \label{deleta} \\
    &\simeq
    \frac{e^{2 \Gamma \tau}}{4 g^2 \tau^2
    (T - \tau)^2} + \frac{e^{2 \Gamma \tau} (\Delta^2 + 3 \sigma^2)
    (2 T^2 - \tau T + \tau^2)}{24 g^2 \tau^2 (T - \tau)^2}
    + \frac{\sigma^2 (2 \bar{n} + 1)}{4} + \frac{\sigma^2 g^2
    \tau^2 (T - \frac{4}{3} \tau)^2}{4 (T - \tau)^2},
\end{align}
in the large $N$ limit.

In the off-resonance case, $\Delta \gg \sigma$, we obtain
\begin{equation}
    B_\delta (T) = g \tau (T - \tau) \left[ {\rm Im} \left( \alpha
    e^{-i \Delta T / 2} \right) \right] \left[ {\rm sinc} \,
    \frac{\Delta (T - \tau)}{2} \, {\rm sinc} \,  \frac{\Delta \tau}{2}
    + {\cal O} (\delta^2 / \Delta^2) \right],
\end{equation}
with ${\rm sinc} \, x = \sin x / x$.
The sensitivity for a single measurement is 
\begin{equation}
    \frac{(\delta \eta)^2}{(\delta \eta)^2 \big\vert_{\Delta = 0}}
    \sim {\rm sinc}^2 \, \frac{\Delta (T - \tau)}{2} \,
    {\rm sinc}^2 \, \frac{\Delta \tau}{2},
\end{equation}
when the measurement time $T$ is short enough to ignore the 
${\cal O} (\sigma^2)$ terms.
For $T = 1$~ms, 5~ms, and 10~ms, the full widths at half maximum are
about 4~kHz, 0.8~kHz, and 0.4~kHz, respectively, corresponding to
the quality factors $Q \sim 10^{3\text{--}4}$, by setting $g = 2 \pi
\times 3.9$~kHz, $\sigma = 2 \pi \times 1$~Hz, $\Gamma = 250$~Hz,
$\bar{n} = 0.3$, and $\tau = 0.1$~ms.

In the main text, 
we use Eqs.~\eqref{16}, \eqref{18}, and \eqref{deleta} to evaluate the sensitivity
$\delta \eta$, assuming that the ODF detuning $\delta$ is normally distributed with a mean of zero and a variance of $\sigma^2$
without expanding over $\delta$.
We then determine the sets of parameters of $T$ and $\tau$ for given experimental parameters
to maximize the signal-to-noise ratio.


\section{Gravitational wave detection with the center-of-mass mode}
\label{sec:com}
In this appendix, we discuss the excitation of the center-of-mass mode of ion crystals for gravitational wave detection.
We first study the direct excitation of the center-of-mass mode
by gravitational waves in Sec.~\ref{orbi}.
The indirect excitation through graviton-photon conversion 
is studied in Sec.~\ref{conv}.


\subsection{Orbiting ion crystal} \label{orbi}

In Sec.~\ref{sec:odd_sensitivity}, we show that only the odd modes respond to gravitational waves,
in contrast to the even modes, when the center of the crystal is fixed.
However, this situation changes if the center is not fixed.
For example, if the center of the crystal exhibits orbital motion,
the parity-even phonon modes including the center-of-mass mode 
which is phase-coherent and has the highest frequency among the phonon
modes, can be excited by gravitational waves. 
We briefly discuss the possibility and estimate the sensitivity of the center-of-mass mode 
to gravitational waves.

Let us assume that an ion crystal undergoes orbital motion.
While this possibility has not yet been explored in existing experiments, such a situation may in principle be realized by applying additional time-dependent electric fields in the crystal plane in addition to the usual experimental setup.
In order to detect external fields, we employ the homogeneous ODF~\eqref{eq:odf}.
At the resonance $\omega_{\rm ODF} = \omega_1 = \omega_z$, the effective
Hamiltonian is written as
\begin{equation}
    \hat{H}_{\rm ODF} = \frac{g}{\sqrt{N}} \left( \hat{a}_1 +
    \hat{a}_1^\dagger \right) \hat{J}_z ,
\end{equation}
under the rotating wave approximation, with $g = F_0 / \sqrt{2 m_{\rm ion}
\omega_z}$.
When we detect external waves, such as electric fields and gravitational
waves, which couple to ions resonantly as
\begin{equation}
    \hat{H}_{\rm int} = \alpha \, \hat{a}_1 + \alpha^* \, \hat{a}_1^\dagger ,
    \label{eq:int_com}
\end{equation}
where $\alpha$ represents the amplitude of the external wave. 
The same protocol as that in Sec.~\ref{sec:odd_protocol} is useful.
The single-measurement sensitivity to $\eta = \vert {\rm Im} \,
\alpha \vert$ is given by Eq.~\eqref{eq:odd_eta0} 
through a parallel discussion for the case of odd mode excitation.

No excitation of the center-of-mass mode by gravitational waves occurs if the center of the ion crystal is fixed, since 
the summation over $x_{i}$ and $y_{i}$ vanishes in Eq.~\eqref{INT}.
However, when the ion crystal exhibits orbital motion,
they are evaluated as
\begin{equation}
    \sum_i x_i =  N r_{\rm orb} \cos (\omega_r t), \quad  
    \sum_i y_i =  N r_{\rm orb} \sin (\omega_r t), \label{orbit}
\end{equation}
where $r_{\rm orb}$
and $\omega_r$ are the radius and angular frequency 
of orbital rotation. 
One obtains the effective Hamiltonian~\eqref{eq:int_com} with
\begin{equation}
    \alpha = - \frac{\sqrt{N m_{\rm ion}} r_{\rm orb} (\omega_z \pm
    \omega_r)^2}{16 \sqrt{\omega_z}} \sin \theta \left[ h^{(+)}
    e^{i \phi^{(+)}} \cos \theta \pm i h^{(\times)} e^{i \phi^{(\times)}}
    \right],
    \label{INT2}
\end{equation}
for $\omega = \omega_z \pm \omega_r$.
We again assume the unpolarized gravitational waves, i.e., $h^{(+)} =
h^{(\times)} = h_0$ and $\phi^{(+)} = \phi^{(\times)} = \phi_0$ for
simplicity. 
Averaging over the polarization and the propagation direction, the signal parameter $\eta = \vert {\rm Im}
\, \alpha \vert$ is given by
\begin{equation}
    \bar{\eta} = \frac{2 \sqrt{2}}{3 \pi} \, K(1/\sqrt{2}) \times
    \frac{\sqrt{N m_{\rm ion}} r_{\rm orb} (\omega_z \pm
    \omega_r)^2}{16 \sqrt{\omega_z}} h_0 .
    \label{eq:etacom}
\end{equation}
We mention that the rotating wave approximation imposes the condition $\omega_r T \gg 1$.
Indeed, the resonant amplification of the phonon mode by a gravitational wave 
becomes significant only after $t \gtrsim \pi / 2 \omega_r$.

Comparing Eq.~\eqref{eq:etacom} to Eq.~\eqref{deleta}, one can estimate the sensitivity to the amplitude of the 
gravitational wave corresponding to ${\rm SNR}=1$ for one-year observation as
\begin{equation}
    h_0(f) = 6.9 \times 10^{-16} \times
    \left( \frac{N}{150} \right)^{-1/2}
    \left( \frac{m_{\rm ion}}{8.3~{\rm GeV}} \right)^{-1/2} 
    \left( \frac{r_{{\rm orb}}}{1~{\rm mm}} \right)^{-1} 
    \left( \frac{f}{1.6~{\rm MHz}} \right)^{-3/2} ,
\end{equation}
which is an order of magnitude superior to that achieved in the odd mode.
Although $r_{\rm orb}$ is assumed to be 1~mm,
ten times longer than the radius of the crystal, it may in principle be increased up to the size of the apparatus, 
in which case a further improvement in sensitivity is expected.


\subsection{Graviton-photon conversion} \label{conv}

It is known that gravitational waves convert into electromagnetic waves in the presence of a background
magnetic field.
In this case, the center-of-mass mode can be excited indirectly
by the converted electromagnetic waves.
Graviton-photon conversion can be described by an
interaction term~\cite{Ratzinger:2024spd},
\begin{align}
    \mathcal{L}_{gp} = j^\mu A_\mu ,
\end{align}
with the effective current 
\begin{align}
    j^\mu = \partial_\nu \left(h^\mu_\lambda \bar{F}^{\lambda\nu}-h^\nu_\lambda\bar{F}^{\lambda\mu}\right) -\frac{1}{2}\left(\partial_\lambda h^\nu_\nu\right)\bar{F}^{\lambda\mu}, 
\end{align}
where $\bar{F}^{\mu\nu}=\partial^\mu\bar{A}^\nu-\partial^\nu\bar{A}^\mu$ is the field strength tensor of the 
electromagnetic waves and $\bar{A}^\mu$ denotes the background magnetic field. 
The interaction induces the electric field under the long-wavelength approximation $\omega L\ll1$ as 
\begin{align}
    E_z \sim \frac{1}{\sqrt{2}}h_0B_z(\omega L)^2\sin^2\theta\cos(\omega t+\phi_0). \label{E}
\end{align}
Here, the background magnetic field $B_z$ is assumed to be cylindrical with radius $L$~\cite{Ouellet:2018nfr}. 
We note that the above expression is valid when boundary effects, such as those imposed by electrodes and 
cavities (if any), are negligible.
From the parallel discussion in Sec.~\ref{orbi}, one can estimate the sensitivity to the amplitude of the 
gravitational wave corresponding to ${\rm SNR}=1$ for one-year observation as
\begin{align}
    h_0(f) \sim 1.7\times10^{-16}\times
    \left( \frac{N}{150} \right)^{-1/2}
    \left( \frac{m_{\rm ion}}{8.3~{\rm GeV}} \right)^{-1/2} 
    \left( \frac{L}{6.5~{\rm cm}} \right)^{-2}
    \left( \frac{B_z}{4.46~{\rm T}} \right)^{-1}
    \left( \frac{f}{1.6~{\rm MHz}} \right)^{-2} .
\end{align}
This indicates that the indirect detection could
be a promising way for probing high-frequency gravitational waves.


\bibliography{bibcollection}
\bibliographystyle{modifiedJHEP}


\end{document}